# Large 1/*f* noise of unipolar resistance switching and its percolating nature

S. B. Lee,[1] S. Park,[2] J. S. Lee,[3] S. C. Chae,[1] S. H. Chang,[1] M. H. Jung,[2] Y. Jo,[2] B. Kahng,[3] B. S. Kang,[4] M.-J. Lee,[4] and T. W. Noh[1,a)]

[1]*ReCOE, Department of Physics and Astronomy, Seoul National University, Seoul 151-747, Republic of Korea*

[2]*Quantum Materials Research Team, Korea Basic Science Institute, Daejeon 305-333, Republic of Korea*

[3]*Department of Physics and Astronomy, Seoul National University, Seoul 151-747, Republic of Korea*

[4]*Samsung Advanced Institute of Technology, Suwon 440-600, Republic of Korea*

---

a)Electronic mail: twnoh@snu.ac.kr

We investigated the 1/$f$ noise of Pt/NiO/Pt capacitors that show unipolar resistance switching. When they were switched from the low to high resistance states, the power spectral density of the voltage fluctuation was increased by approximately five orders of magnitude. At 100 K, the relative resistance fluctuation, $S_R/R^2$, in the low resistance state displayed a power law dependence on the resistance $R$: i.e., $S_R/R^2 \propto R^w$, where $w = 1.6 \pm 0.2$. This behavior can be explained by percolation theory; however, at higher temperatures or near the switching voltage, $S_R/R^2$ becomes enhanced further. This large 1/$f$ noise can be therefore an important problem in the development of resistance random access memory devices.

Recently, unipolar resistance switching (RS) phenomena has attracted much attention, due to potential applications in resistance random access memory (RRAM).[1–4] From an application viewpoint, the signal-to-noise ratio of RRAM should be as large as possible. Additionally, measurements of noise can provide valuable information on the electronic transport and the microscopic switching mechanisms that cannot be acquired by other measurements.[5–7] Despite its scientific and technological importance, there has been little investigation of the noise behavior of unipolar RS.

Here, we report on $1/f$ noise of unipolar RS by using Pt/NiO/Pt capacitors. When we applied a bias, $V$, to our samples, we observed that the power spectral density, $S_V$, of the low frequency voltage fluctuation was inversely proportional to frequency, $f$. We found that the $1/f$ noise in the low resistance state (LRS) increased significantly as the corresponding resistance increased. Detailed behavior of the $1/f$ noise will be discussed within the framework of the percolation theory. We will also present some experimental data that cannot be understood by such classical theory.

We deposited polycrystalline NiO thin films on Pt-coated Si substrates, using DC magnetron reactive sputtering. To measure the electrical properties and noise, we fabricated Pt/NiO/Pt capacitors by depositing Pt top electrodes with an area of 50×50 $\mu$m$^2$ onto the NiO layer. Figure 1(a) shows current–voltage ($I$–$V$) curves of a Pt/NiO/Pt capacitor, which exhibits unipolar RS. The NiO capacitor was highly insulating in the pristine state (PS), marked by the black circles. Immediately after the forming process at $V_F$, it enters the LRS. When $V$ was increased above $V_R$, it switched from the LRS to the high resistance state (HRS). This switching behavior was termed a reset process. As we increased $V$ further, reaching $V_S$, the capacitor switched back into the LRS. This was termed a set process. To prevent permanent damage (i.e., complete dielectric breakdown) during both forming and set processes, we used a compliance current $I_{comp}$ of 1 mA.

We applied a bias to the Pt/NiO/Pt capacitor using a source-meter unit (Keithley 2601) and measured the noise intensity, $S_V$, using a spectrum analyzer (Agilent E4448A).[8] To amplify the $S_V$ signal, two kinds of low-noise differential amplifiers were used (NF SA-400F3 for LRS, and ITHACA 1201 for HRS and PS). We also used a Wheatstone bridge, in which one of the resistors was replaced with our Pt/NiO/Pt capacitor, to measure the small noise signal with a high sensitivity. As shown in Fig. 1(c), we observed that the $S_V$ signals in all of the resistance states had $1/f^\alpha$ dependence under an external bias. Since $\alpha \approx 1$, we will refer to the $S_V$ signal as "$1/f$ noise" from now on.

Even when $V = 0$ V, there should be $f$-independent voltage fluctuations $S_V^{th}$ (i.e.,

Johnson noise[5,7]). According to the fluctuation-dissipation theorem, the corresponding $S_V^{th}$ should be proportional to the resistance, $R$, and temperature, $T$. The theoretical values of $S_V^{th}$ for the PS and HRS are shown by the dashed lines in Fig. 1(c). Our experimental $S_V^{th}$ for the PS and HRS are in good agreement with the theoretical predictions. The theoretical $S_V^{th}$ for the LRS should be smaller, by five orders of magnitude, than that for the HRS; however, this was below the experimental sensitivity. We also investigated the $V$-dependence of $S_V$ in the LRS. In the Ohmic conduction region for most metals, $S_V$ should follow Hooge's equation (i.e., $S_V \propto V^2$).[5–7] As shown by the blue line in Fig. 1(b), $S_V \propto V^2$ in the LRS with $V < 0.15$ V, where Ohmic conduction occurs. These confirmations indicate that our 1/$f$ noise measurements were highly reliable.

Figure 1(b) shows the $S_V$-values of a Pt/NiO/Pt capacitor, measured at room temperature with $f = 100$ Hz. For any resistance state, as $V$ increases, $S_V$ increases slowly when $V$ is small, but experiences a rapid increase near the corresponding switching voltage. We can readily convert $S_V$ to the power spectral density of the resistance fluctuation, $S_R$, using $S_R = S_V/I^2$ (Ref. 7). While $R^{HRS}$ was larger than $R^{LRS}$ by five orders of magnitude, $S_R^{HRS}$ was found to be larger than $S_R^{LRS}$ by 15 orders of magnitude at 0.2 V.

Why does the 1/$f$ noise become so large in the HRS? It is widely accepted that unipolar RS can occur due to the formation and rupture of conducting filaments under an applied electric field.[1–4,9–13] The LRS is characterized by the conducting filaments making percolating channels across the sample. When the channels become broken, by Joule heating, the HRS applies. 1/$f$ noises in percolating material systems have been treated by classical percolation theories.[7,8,14] It is known that $S_R$ probes the fourth moments of current distributions in the percolating network.[7,14] More explicitly, $S_R/R^2$ can be written as

$$\frac{S_R}{R^2} = \frac{\sum \rho_b^2 i_b^4}{\left(\sum r_b i_b^2\right)^2}, \qquad (1)$$

where $i_b$ is the current flowing through each filament $b$, whose resistance value is $r_b + \delta r_b$, and $\rho_b^2 = \langle \delta r_b \delta r_b \rangle$ is determined by the mechanism of fluctuations in the filament, which typically varies as the 1/$f$ type. As the fraction, $p$, of conducting filaments approaches to the percolation threshold, $p_c$, $S_R$ should diverge as $S_R \propto (p - p_c)^{-\kappa}$, where $\kappa$ is the noise exponent.[7,8] This divergence near $p_c$ might explain the drastic increase of $S_R$ in the HRS.

We performed quantitative studies on the relative noise $S_R/R^2$. According to percolation theory,[7,8,14]

$$\frac{S_R}{R^2} \propto R^w \text{ at } p > p_c. \qquad (2)$$

To measure $S_R$ for numerous resistance values in the LRS near $p_c$, we generated some multilevel LRS by carefully increasing $V$ near $V_R$ (Fig. 2(a)).[15] Similarly, several multilevel HRS were also generated (Fig. 2(b)). At each multilevel state, we measured $S_V$ at room temperature with $V = 0.1$ V, and plotted the corresponding $S_R/R^2$-value in Fig. 2(c). In the LRS, $S_R/R^2$ has the power law dependence with $w = 1.8 \pm 0.3$. In the HRS, $S_R/R^2$ becomes very large and scattered around $10^{-8}$ Hz$^{-1}$.

To obtain further insight, we measured $S_R$ in the LRS by varying $T$ and $V$. As shown in Fig. 3(a), with a decrease of $T$ from 300 K to 100 K, and with $V = 0.1$ V, $w$ decreased, from 1.8 to 1.6. However, as shown in Fig. 3(c), with $V = 0.3$ V, $w$ decreased drastically, from 5.1 to 1.6. It should be noted that all of the values of $w$ at 100 K were nearly the same, i.e., $w = 1.6 \pm 0.2$, regardless of $V$.

Note that percolation theory deals with the enhancement of $S_R/R^2$ due to the current distribution in the percolating network.[7,8,14] Specifically, it deals with a purely geometrical effect, assuming that the resistance fluctuation in the conducting filament networks should be the same. So, the observation that $w = 1.6 \pm 0.2$ at low $T$ may be related to a geometric effect. According to classical lattice percolation theory,[7,8] $0.82 < w < 1.05$. For two-dimensional (2D) semicontinuous metal films,[7,8] it has been reported that $1.2 < w < 2.0$. To explain the unipolar RS phenomena, we recently developed a new type of percolation model, termed 'the random circuit breaker (RCB) network model,' where fuse and antifuse actions occur, depending on the voltage applied to a filament.[11] We performed 2D numerical simulations using the RCB network model and found $w = 1.5 \pm 0.3$. This value agrees quite well with our observed $w$ at 100 K, suggesting that the low temperature $S_R/R^2$ should be related to the current distribution in the percolating network.

The drastic increase of $S_R/R^2$ in the LRS near the reset voltage $V_R$ (especially at high $T$) should be due to additional effects. The first possibility may be Joule heating effects, especially near the weakest bond, where most of the current will flow.[1,3,4,9,13,15] To investigate this, we included Joule heating and associated thermal dissipation processes in the RCB network model.[11,13,16] However, we still obtained $w = 1.5 \pm 0.3$. Another possibility is motion of ions near the resistance switching, which are currently discussed by many workers.[1,3,4] Further systematic investigations are needed to investigate this phenomenon.

A further important point related to the noises in the unipolar RS is that at $V = V_S$, $S_V^{HRS} \approx 10^{-4}$ V$^2$/Hz (Fig. 1(b)). This is much larger than the $S_V$-value of phase-change random access memory[17] (where $S_V \approx 10^{-10}$ V$^2$/Hz), and that of magnetic random access memory[18] (where $S_V \approx 10^{-15}$ V$^2$/Hz). For RRAM applications, the large noise issue should be carefully addressed.

In summary, we performed systematic investigations into the 1/$f$ noise of unipolar resistance switching. Our experimental 1/$f$ noise data at 100 K and low applied bias confirms our view that unipolar resistance switching should occur by the percolation process of conducting filaments. However, at higher temperature, or near the switching voltages, 1/$f$ noise became further enhanced, indicating the existence of additional noise sources. Our work indicates that 1/$f$ noise of unipolar resistance switching is an important issue and should be investigated further for RRAM device applications.

This work was supported by the Korea Science and Engineering Foundation (KOSEF) grant, funded by the Korea government (MEST) (No. 2009-0080567). B.K. and J.S.L. were supported by the KOSEF grant funded by the MOST (Grant No. R17-2007-073-01001-0).

FIG. 1. (Color online) (a) *I–V* characteristics of a Pt/NiO/Pt capacitor. The LRS shows Ohmic behavior ($I \propto V$, blue line) below $V_o = 0.15$ V. (b) *V*-dependence of $S_V$. At the Ohmic region, $S_V^{LRS}$ follows Hooge's equation, i.e., $S_V \propto V^2$. As *V* approaches $V_R$, $V_S$, and $V_F$, respectively, $S_V$ starts to increase rapidly. (c) *f*-dependence of $S_V$. With $V = 0$ V, $S_V^{th}$ of each resistance state (open symbols) follows theoretical values of Johnson noise (the dashed lines). With $V > 0$ V, each state shows $1/f^\alpha$ noise (solid symbols), with $\alpha$ close to unity. Note that $S_V^{PS} > S_V^{HRS} > S_V^{LRS}$ under $V = 0$ V and $S_V^{HRS} > S_V^{PS} > S_V^{LRS}$ under $V > 0$ V.

FIG. 2. (Color online) Obtaining multilevels of (a) the LRS and (b) the HRS by carefully controlling *V*. (c) The *R*-dependence of $S_R/R^2$ in the LRS and HRS. While $S_R/R^2$-values of the LRS are proportional to $R^w$ with $w = 1.8 \pm 0.3$, the HRS shows very large $S_R/R^2$, scattered around $10^{-8}$ Hz$^{-1}$.

FIG. 3. (Color online) The *R*-dependence of $S_R/R^2$ in the LRS for $V =$ (a) 0.1 V, (b) 0.2 V, and (c) 0.3 V. The red, black, and blue triangles show $S_R/R^2$ at 100 K, 200 K, and 300 K, respectively. Note that *w* increases with increasing *V* and *T*.

*Figure 1*

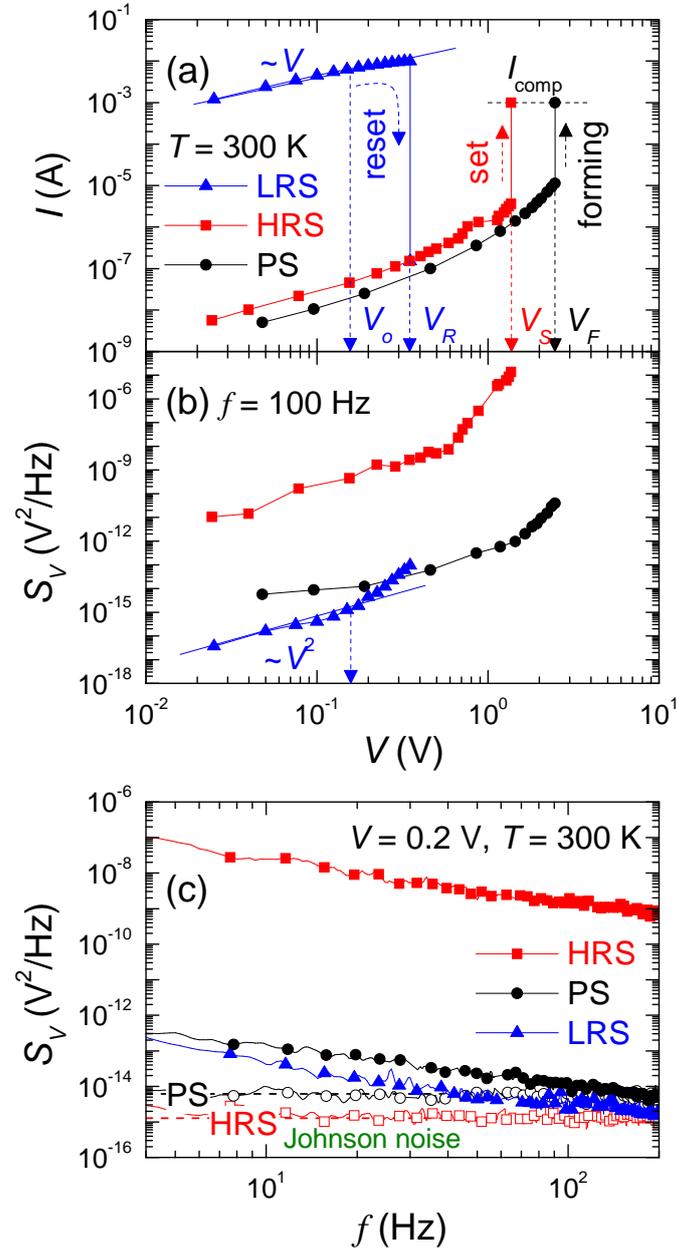

*Figure 2*

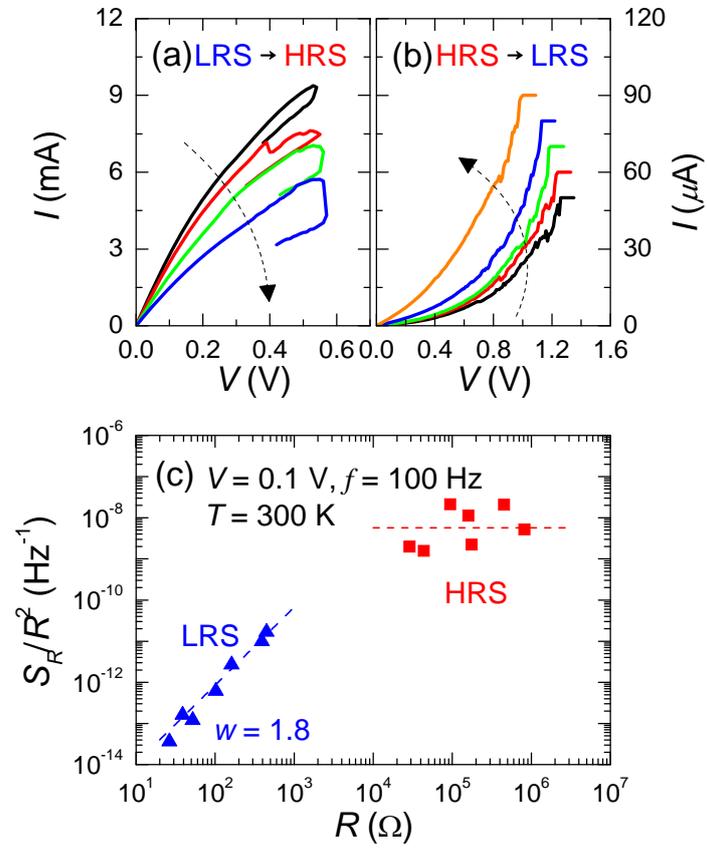

*Figure 3*

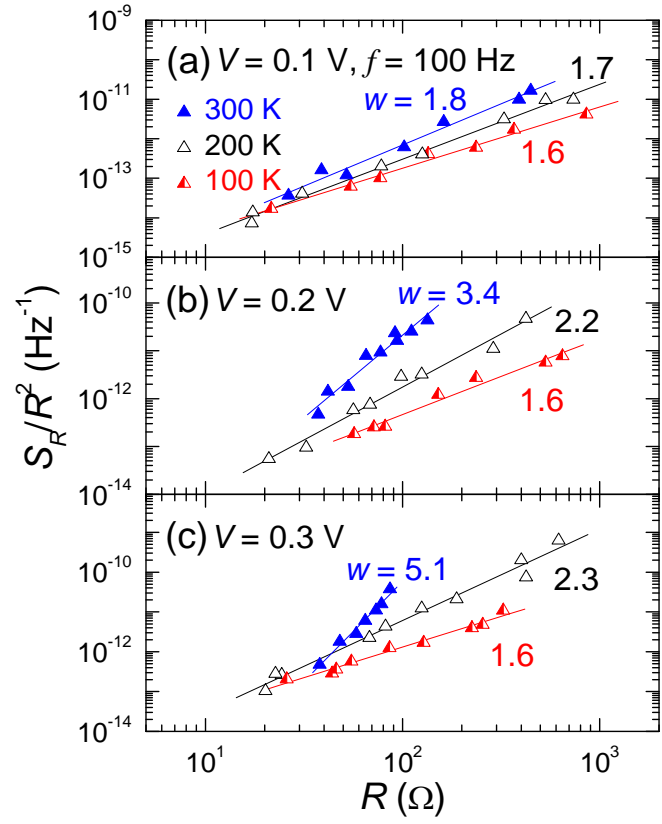